\documentstyle[12pt]{article}
\advance\textheight by \topskip
\begin{document}
\def\odd{$O(d,d)$}
\def\oddh{$O(d,d+16)$}
\def\be{\begin{equation}}
\def\ee{\end{equation}}
\def\beq{\begin{eqnarray}}
\def\eeq{\end{eqnarray}}
\def\wt{\widetilde}
\def\del{\partial}
\def\s{$\sigma$}
\renewcommand{\thefootnote}{\fnsymbol{footnote}}
\begin{center}
\hfill\\\hfill\\
{\Large\bf T-Duality and Non-Local Supersymmetries}\\
\vskip 1.5 cm
{\bf S. F. Hassan}\footnote{ E-mail address:
{\tt fawad@surya1.cern.ch}}
\vskip 0.1cm
{\it Theory Division, CERN \\ CH-1211 Geneva 23, Switzerland}\\
\end{center}
\vskip 2cm
\centerline{\bf ABSTRACT}
\begin{quotation}
\noindent
We study the non-localization of extended worldsheet supersymmetry
under T-duality, when the associated complex structure depends on the
coordinate with respect to which duality is performed. First, the
canonical transformation which implements T-duality is generalized to
the supersymmetric non-linear $\sigma$-models. Then, we obtain the
non-local object which replaces the complex structure in the dual
theory and write down the condition it should satisfy so that the dual
action is invariant under the non-local supersymmetry. For the target
space, this implies that the supersymmetry transformation parameter is
a non-local spinor. The analogue of the Killing equation for this
non-local spinor is obtained. It is argued that in the target space,
the supersymmetry is no longer realized in the standard way.
The string theoretic origin of this phenomenon is briefly discussed.
\end{quotation}
\vspace{1cm}
\begin{flushleft}
CERN-TH/95-98\\
hep-th/9504148\\
April 1995
\end{flushleft}
\newpage
\renewcommand{\thefootnote}{\arabic{footnote}}
\setcounter{footnote}{0}

\section{Introduction}

It is known that one should expect certain non-local effects to appear
in an effective field theory based on string theory \cite{GSW}.  This
clearly has to do with the fact that, unlike a point particle, the
string is not a dimensionless object. Such effects in the low-energy
theory have not yet been studied in detail, probably due to the
absence of compulsive evidence for their importance.  However, there
is significant evidence that they do appear in connection with
important issues such as T-duality and supersymmetry
\cite{K2,AAL1,AAL2,BS}.  In this paper we will report on an
investigation of the issue of non-localization of extended worldsheet
supersymmetry and the associated target space supersymmetry under a
T-duality transformation.

A given conformal field theory may have different target space
realizations which are related to each other by a T-duality
transformation \cite{Busc,K1,RV,GR,Review1,Review2}. The mechanism by
which a T-duality transformation gives rise to a non-locality (in the
target-space sense) is most transparent when the duality
transformations are formulated as canonical transformations in the
worldsheet theory \cite{AAL2,GRV,MV,KS}. In this approach, the
coordinate (say $\theta$) with respect to which duality is performed
and the corresponding coordinate in the dual theory (say
$\widetilde\theta$) are non-local functions of each other. The
non-locality is a consequence of an integration over the string length
parameter which appears in the relation between $\theta$ and
$\wt\theta$. As a result, any $\theta$-dependent quantity in one
theory becomes a non-local function of the corresponding coordinate in
the dual theory. This effect, which is also accompanied by the
interchange between momentum modes ({\it local}) and winding modes
({\it non-local}), is solely due the extended nature of the string.
Since a duality transformation with respect to the coordinate $\theta$
is performed only when the massless background fields are independent
of $\theta$, the non-localities will not show up unless we go beyond
this set of fields. An example is a WZNW model based on a group $G$
and the corresponding symmetry currents. In this case, the group $G$
may have a non-local realization in the dual theory
\cite{K2,AAL1,AAL2}.

A more interesting situation, in which non-local effects show up after
a duality transformation, arises when the worldsheet theory has an
extended supersymmetry \cite{ZAFM,GHR,HW,HP,SenN=2}. It is known that
if a complex structure associated with an extended supersymmetry on
the worldsheet, does not have a dependence on the coordinate $\theta$,
then in the dual model the extended supersymmetry is realized in the
usual way \cite{KIR, H}. However, it was noticed in \cite{Bakas} that in
certain examples supersymmetry is not preserved under a duality
transformation. It turns out that in all these examples, the complex
structure associated with the supersymmetry under consideration
depends on the coordinate $\theta$ with respect to which duality is
performed. Then, from the discussion above, it follows that in the
dual theory the complex structure is replaced by a non-local object.
A prescription for obtaining this non-local object was suggested in
\cite{BS}. This phenomenon implies that, in such a situation, the
extended supersymmetry of the dual theory, though still present, is
realized non-locally.  In particular, the relation between
supersymmetry and target space geometry is modified. Since worldsheet
supersymmetry is intimately connected with target space supersymmetry,
this non-locality is also expected to have implications for the
latter. However, one should keep in mind that as string theories, the
dual models are, nevertheless, physically equivalent.

The mechanism by which the non-locality arises is not specific to
duality transformations. It is, in fact, common to all non-trivial
$O(d,d)$ transformations which generically relate physically
inequivalent background field configurations and which contain
T-duality as a discrete subgroup.

In this paper, we investigate the phenomenon of non-localization of
supersymmetry under a T-duality transformation.  The paper is
organized as follows: In section 2, we describe our conventions and
then generalize the canonical transformation which implements
T-duality from bosonic to supersymmetric non-linear \s-models. This
canonical transformation can be written in the superfield notation.
In section 3, we consider a non-linear \s-model with extended
supersymmetry on the worldsheet such that the associated complex
structure is $\theta$-dependent. We obtain the non-local object which,
in the dual theory, replaces the $\theta$-dependent complex structure.
We then obtain the conditions on this non-local object which are
analogous to the covariant constancy of complex structure in the
standard realization of the extended supersymmetry. In section 4, we
show that, in the dual theory, the Killing spinor associated with the
target space supersymmetry is also replaced by a non-local spinor.  We
obtain the analogue of the Killing equation which this non-local
spinor satisfies and discuss its consequence for the realization of
non-local target space supersymmetry.  At the end, we briefly discuss
the string theoretic origin of the non-locality when $\theta$ is a
compact coordinate. In section 5, we summarize our results and point
out an  example where S-duality seems to be incompatible with the
standard realization of worldsheet supersymmetry.

\section{T-Duality as a Canonical Transformation in Supersymmetric
Theories }

In this section, we generalize the method of realizing a T-duality
transformation by a canonical transformation on the worldsheet to
include $N=1$ supersymmetric non-linear \s-models.

We start with the non-linear \s-model with $N=1$ supersymmetry on the
worldsheet. Following the conventions of \cite{STPS}, in the component
notation the action defining this model takes the form
\beq
\label{action}
S&=&{1\over 2}\int d^2 \sigma \big[ (G_{MN}+B_{MN})\del_+ X^M \del_-X^N
\nonumber \\
&-&i \psi_+^M G_{MN}(\delta_K^N\del_- + \Omega^{+N}_{LK}\del_-X^L)\psi_+^K
-i \psi_-^M G_{MN}(\delta_K^N\del_+ + \Omega^{-N}_{LK}\del_+X^L)\psi_-^K
\nonumber \\
&+&{1\over 2} \psi_+^M \psi_+^N \psi_-^K \psi_-^L R_{MNKL}(\Omega^-)
\big].
\eeq
Here, $\Omega^{\pm}$ are the torsionful connections given by
$\Omega^{\pm K}_{MN}=\Gamma^K_{MN} \pm {1\over 2} G^{KL}H_{LMN}$, where
$\Gamma^K_{MN}$ is the Christoffel symbol and $H_{MNK}$ is the torsion
tensor given by $H_{MNK}= 3\del_{[M} B_{NK]}$; $R_{MNKL}(\Omega^\pm)$
are the curvature tensors corresponding to the torsionful connections
and satisfy the property $R_{MNKL}(\Omega^-)=R_{KLMN}(\Omega^+)$. The
above action has a default $N=1$ supersymmetry which has independent
action on the left-moving and the right-moving chiral sectors of the
theory. The transformations of the fields under this $(1,1)$
supersymmetry are given by
\beq
\label{N=11}
\delta_\mp X^M &=& \pm i \epsilon_\mp\psi_\pm^M
\\
\label{N=12}
\delta_\mp\psi_\pm^M &=&\pm\del_\pm X^M\epsilon_\mp
\\
\label{N=13}
\delta_\mp\psi_\mp^M &=& \mp i\psi_\mp^N\epsilon_\mp
\Omega^{\pm M}_{NK} \psi_\pm^K .
\eeq
In the following, we describe the implementation of T-duality
transformations by canonical transformations in the above $N=1$
theory. Similar issues in the context of the chiral model have been
discussed in \cite{CZ}. In different contexts, T-duality in
supersymmetric theories has been studied in \cite{also}.

A T-duality transformation is always performed with respect to a
Killing vector field $K$ defined by ${\cal L}_K G=0$, ${\cal L}_K
B=d\omega_{(K)}$ and ${\cal L}_K \Phi =0$. Here, ${\cal L}_K$ denotes
the Lie derivative along the vector field $K$ and $\omega_{(K)}$ is a
one-form in the target space; $\Phi$ is the dilaton field, which does
not appear in (\ref{action}). We choose a coordinate system
$ X^1=\theta, X^{i+1}=x^i; i= 1,...,D-1$ such that the Killing vector
takes the form $K= \del/{\del\theta}$. In this coordinate system, the
background fields $G$ and $B$ can be chosen to be independent of
$\theta$ and, under a duality transformation with respect to $\theta$,
transform to
\beq
&&\wt G_{\theta \theta} = G^{-1}_{\theta\theta},
\nonumber \\
&&(\wt G+\wt B)_{\theta i}=-G^{-1}_{\theta\theta}(G+B)_{\theta i},
\nonumber\\
&&(\wt G-\wt B)_{\theta i}=G^{-1}_{\theta\theta}(G-B)_{\theta i},
\nonumber \\
\label{dual}
&&(\wt G+\wt B)_{ij}=(G+B)_{ij}-
G^{-1}_{\theta\theta}(G-B)_{\theta i}(G+B)_{\theta j} .
\eeq
In many situations, as in equations (\ref{N=11})--(\ref{N=13}), the
field $B_{MN}$ appears only in the torsionful connections $\Omega^\pm$
through its field strength tensor $H_{MNK}$. In these situations, it
is convenient to rewrite the duality transformations in terms of the
relevant variables which are $G_{MN}$ and $\Omega^{\pm M}_{NK}$. To do
this, let us introduce two $D\times D$ dimensional matrices $Q_\pm$
given by \footnote{The matrices $Q_\pm$ were introduced in \cite{H},
  in   connection with arbitrary non-trivial $O(d,d)$
  transformations. For the discrete duality subgroup, they reduce to
  the ones given above.}
\be
\label{Qs}
Q_\pm =\left(
\begin{array}{cc}\mp G_{\theta\theta} & \mp(G\mp B)_{\theta i} \\
                      0             &  1_{D-1}
\end{array}\right)\,,
\qquad
Q_\pm^{-1} = \left(
\begin{array}{cc}
\mp G^{-1}_{\theta\theta} & -G^{-1}_{\theta\theta}(G\mp B)_{\theta i} \\
                      0             &  1_{D-1}
\end{array}\right) .
\ee
In terms of these, the dual metric is given by
\be
\label{dmetric}
\wt G^{-1} = Q_- G^{-1} Q_-^T = Q_+ G^{-1} Q_+^T \,,
\ee
and the dual torsionful connections take the form
\be
\label{dconnec}
\wt\Omega^{\pm M}_{NK}=(Q^{-1}_\mp)^{N'}_{~N}\,(Q^{-1}_\pm)^{K'}_{~K}
\,(Q_\mp)^M_{~M'}\, \Omega^{\pm M'}_{N'K'}
-\delta^i_K\,(\del_i\,Q_\mp\,Q^{-1}_\mp)^M_{~N} .
\ee
The above equation can be obtained using the formalism in \cite{H}.
Note that though the metric transforms unambiguously, the (inverse)
vielbein $e$, defined by $G^{-1}=e\eta e^{T}$, does not. However, the
two possible options, $\wt e_+=Q_+e$ and $\wt e_-=Q_-e$, are related
by a local Lorentz transformation: $\wt e_+ = \wt e_-\Lambda$. Here,
$\Lambda$ is given by $\Lambda =e^{-1}Q_-^{-1}Q_+e$ and it
satisfies $\Lambda \eta\Lambda^T=\eta$. Note that, in the above, we
have assumed $e$ to be $\theta$-independent.

The transformations of the background fields (\ref{dual}) can be
obtained in different ways depending on the method one chooses to
implement duality.  However, the ease with which these methods could
be generalized to produce the transformation under duality of other
operators in the theory, varies from method to method. From this point
of view, the canonical approach seems to be the most powerful.  In
bosonic non-linear \s-models, the implementation of duality by a
canonical transformation was first used in \cite{GRV} for constant
background fields and then discussed in more detail in \cite{AAL2}:
Let $p_\theta$ denote the canonical momentum conjugate to the
coordinate $\theta$ and let $\theta'=\del\theta/\del\sigma$. The
duality transformations (\ref{dual}) now follow from the canonical
transformation
\be
\label{ct-b}
\wt \theta'=-p_\theta\,, \qquad \wt p_\theta = - \theta'\,,
\qquad\wt x^i = x^i \,.
\ee
It is clear from the above that the relation between $\theta$ and
$\wt\theta$ is, in general, non-local.

We want to apply this procedure to the supersymmetric model defined by
the action (\ref{action}). The $\theta$-independence of the background
fields in this model gives rise to a conserved isometry current with
components ${\cal I}_\pm$ given by
\be
\label{isometry}
{\cal I}_\pm={1\over 2}(G\mp B)_{\theta M}\,\del_\pm X^M-{i\over 2}
\,F_\pm
\ee
where we have used the notation
\be
\label{Fs}
F_\pm\equiv\psi_\pm^M\,G_{MN}\,\Omega^{\pm N}_{\theta L}\,\psi_\pm^L
     =\psi_\pm^M\,\psi_\pm^j\,\del_j (G\mp B)_{\theta M} .
\ee
The canonical momentum conjugate to $\theta$ can be obtained from
(\ref{action}) and is given by
\beq
p_\theta
&=&{\cal I}_+ +{\cal I}_-
\nonumber\\
\label{momen}
&=& G_{\theta \theta}\dot\theta+{1\over 2}(G-B)_{\theta j}\,\del_+ x^j
+{1\over 2}(G+B)_{\theta j}\,\del_- x^j-{i\over 2}(F_++F_-) .
\eeq
It is clear that in order to obtain the dual theory in the
supersymmetric case, the canonical transformation (\ref{ct-b}), with
$p_\theta$ given by (\ref{momen}), is not sufficient. It has to be
accompanied by the appropriate transformations of the worldsheet
fermions so that $N=1$ supersymmetry is preserved. The required
transformations of fermions can be obtained by demanding that the
supersymmetry transformation equations (\ref{N=11})--(\ref{N=13})
imply a similar set of equations for the dual theory, provided the
backgrounds in the two theories
are still related by (\ref{dual}). First, consider eq.
(\ref{N=12}) which, in the canonically transformed theory, should take
the form $\delta_\mp\wt\psi_\pm^M=\del_\pm\wt X^M\epsilon_\mp$. Since the
canonical transformation (\ref{ct-b}) does not affect the coordinates
$x^i$, we get $\wt\psi_+^i=\psi_+^i$. For the $\psi^\theta$-component,
using (\ref{ct-b}) and (\ref{momen}) along with the duality relations
(\ref{dual}), we obtain $\wt\psi_\pm^\theta=
\mp(G\mp B)_{\theta M}\psi_\pm^M$. These transformations of fermions,
which should accompany the canonical transformation
(\ref{ct-b}), can be written in terms of the matrices $Q_\pm$ as
\be
\label{ct-f}
\wt\psi_\pm^M = Q^M_{\pm N}\psi_\pm^N .
\ee
The origin of the difference in transformations of $\psi^M_+$ and
$\psi^M_-$ can be traced back to the interpretation of duality (and
for that matter, all non-trivial $O(d,d)$ transformations) as Lorentz
transformations acting independently on the two chiral sectors of the
underlying conformal field theory \cite{twist}. Note that since the
zero modes of worldsheet fermions are related to the Clifford algebra
in the target space, the above difference may have an implication for
the target space supersymmetry. However, as will be discussed in
section 4, this difference can be absorbed in a redefinition of the
corresponding target space spinors. The compatibility of the
transformations (\ref{ct-b}) and (\ref{ct-f}) with the remaining two
of the $N=1$ supersymmetry transformation equations (\ref{N=11}) and
(\ref{N=13}) will be discussed below.

Let us consider the relation between the original and the dual
coordinates. Using (\ref{ct-b}) and (\ref{ct-f}), this can be written as
\be
\label{ct-bp}
\del_\pm\wt\theta=Q^\theta_{\pm M}\,\del_\pm X^M +i\psi^j_\pm\,\del_j
Q^\theta_{\pm M}\,\psi^M_\pm .
\ee
Comparing the above equations with (\ref{isometry}), we get
$\del_\pm\wt\theta =\mp 2{\cal I}_\pm$, so that
\be
\label{below}
\wt\theta=2\int d\sigma^-{\cal I}_--2\int d\sigma^+{\cal I}_+ .
\ee
The above relation tells us that the dual coordinate is a non-local
function of the original coordinates. However, the conservation of the
isometry current, $\del_+{\cal I}_- +\del_-{\cal I}_+=0$, leads to
$\del_+\del_-\wt\theta=\del_-\del_+\wt\theta$. This implies that
in spite of the non-local relation between $\theta$ and $\wt\theta$,
on shell, the dual coordinate is well defined on the worldsheet and is
a local function of the worldsheet coordinates. As a digression,
notice that  if $G_{\theta \theta}=1$ and $(G+B)_{\theta i}=0$, or
equivalently if $Q_-=1_D$, then the backgrounds are self-dual and we
do not expect a non-locality to show up in the transformations. In
fact in this case, $\del_-\wt\theta=\del_-\theta$ which has a solution
$\wt\theta=\theta+f(\sigma^+)$, where the function
$f(\sigma^+)$ is determined by $\del_+\wt\theta=-2{\cal I}_+$.
However, the isometry current conservation equation now takes the form
$\del_-(\del_+\theta +2 {\cal I}_+)=0$. This defines a chirally
conserved current and the action (\ref{action}) develops an
invariance under $\sigma^+$-dependent translations of $\theta$. This
invariance can be used to eliminate $f(\sigma^+)$ and set
$\theta=\wt\theta$. This proves that a duality transformation with
respect to a chiral isometry does not result in a non-locality
\cite{AAL1,AAL2}.

Now, we consider the compatibility of the canonical transformations
(\ref{ct-b}) and (\ref{ct-f}) (or equivalently, (\ref{ct-f}) and
(\ref{ct-bp})) with the remaining two of the $N=1$ supersymmetry
transformations, {\it i.e.} eqs. (\ref{N=11}) and (\ref{N=13}).
First, using (\ref{dconnec}), it is easy to see that under a canonical
transformation, (\ref{N=13}) goes over to a similar equation in the
dual theory and is therefore compatible with our transformations. As
for eq. (\ref{N=11}), only the $\theta$-component is non-trivial.
Due to the non-local relation between $\theta$ and $\wt\theta$, one
way to study the behaviour of this equation under a canonical
transformation is to consider its derivatives with respect to
$\sigma^\pm$. This is sufficient because the action is already
invariant under constant shifts of $\theta$. For the sake of clarity, let
us concentrate on the equation involving $\epsilon_-$. Here, it is
again easy to show that the derivative of (\ref{N=11}) with respect to
$\sigma^+$ gives rise to a similar equation for the dual variables,
{\it i.e.} $\delta_-(\del_+\wt\theta)=\epsilon_-\del_+
\wt\psi^\theta_+$. The $\sigma^-$ derivative of (\ref{N=11}) is
slightly different and after a canonical transformation leads to
\be
\delta_-(\del_-\wt\theta) = i\epsilon_- (\del_-\wt\psi^\theta_+)
-4 \epsilon_- {\delta S\over \delta \psi^\theta_+} .
\ee
On shell, $\delta S/ \delta \psi^\theta_+ =0$, and the above equation
reduces to the desired supersymmetry transformation for the dual
theory. This completes the proof that the canonical transformation
(\ref{ct-b}) accompanied by the transformations (\ref{ct-f}) of the
worldsheet fermions (or equivalently transformations (\ref{ct-f}) and
(\ref{ct-bp})) are compatible with the $N=1$ supersymmetry
transformations (\ref{N=11})--(\ref{N=13}).

The canonical transformations (\ref{ct-bp}) and (\ref{ct-f}) can be
written, in a compact form, in terms of the $N=1$ superfields $\Phi^M$
as
\be
\label{superfield}
D_\pm\,\wt\Phi^M = Q^M_{\pm N}(\Phi)\,D_\pm\,\Phi^N .
\ee
The above equation holds on shell and also contains the transformation
of the auxiliary field under duality which is consistent with its
equation of motion.

Since the fermion couplings in (\ref{action}) are entirely determined
by the $N=1$ supersymmetry, the above discussion guarantees that the
canonically transformed action has the same form as the original
action (\ref{action}) with the backgrounds $G_{MN}, B_{MN}$ replaced
by their dual counterparts given by (\ref{dual}). This can be checked
by going to the Hamiltonian formulation where the canonical
transformation takes the form $\wt H(\wt
p_\theta,\wt\theta',\wt\psi_\pm)= H(p_\theta,\theta',\psi_\pm)$.
Rewritten back in terms of the Lagrangian, this gives
\be
\wt L(\del_\pm\wt\theta,\wt\psi^\theta_\pm)=
L(\del_\pm\theta,\psi^\theta_\pm)+ {1\over
  2}(\del_+\wt\theta\,\del_-\theta - \del_-\wt\theta\,\del_+\theta)
\ee
where the variables $\del_\pm\theta, \psi^\theta_\pm$ on the
right-hand side have to be
expressed in terms of $\del_\pm\wt\theta, \wt\psi^\theta_\pm$ using
(\ref{ct-bp}) and (\ref{ct-f}). Now it is a matter of calculation to
check that the Lagrangians $L$ and $\wt L$ have exactly the same form
when the background fields $G$ and $B$ appearing in the two are
related by (\ref{dual}). Going through this calculation, one can see
that the fermion-dependent terms $F_\pm$ appearing in the expression
for the canonical momentum $p_\theta$ contribute only to the
four-fermion terms in the dual theory. Their presence is therefore
necessary to reproduce the correct transformation of the curvature
tensor under duality. By comparing $L$ and $\wt L$, we can very easily
obtain the following transformation equation involving the curvature
tensor, which we note down for
later use:
\be
\label{dualcurv}
\wt\psi^M_+\wt\psi^N_+\wt\psi^K_-\wt\psi^L_-\wt R_{MNKL}(\wt\Omega^-)
=\psi^M_+\psi^N_+\psi^K_-\psi^L_- R_{MNKL}(\Omega^-)
+2 G^{-1}_{\theta\theta}F_+F_- .
\ee
Having generalized the canonical approach to T-duality to the case of
supersymmetric non-linear \s-models, in the next section we turn to
the issue of its effect on extended worldsheet supersymmetry.

\section{T-Duality and Non-Local Extended Supersymmetry}

In this section, we consider a \s-model with extended supersymmetry on
the worldsheet such that the complex structure associated with the
extended supersymmetry is not independent of the coordinate with
respect to which duality is performed. We obtain the non-local object
which replaces the complex structure in the dual theory and write down
the equation which it should satisfy so that the dual action is
invariant under the corresponding non-local supersymmetry
transformations.

We begin with a review of the usual realization of the extended
worldsheet supersymmetry in order to facilitate comparison with its
non-local realization in the dual theory.  If the target space
manifold admits almost complex structures $J^M_{\pm N}$
($J_\pm^2=-1$), then one can obtain a second set of supersymmetry
transformations for $X^M$ and $\psi^M_\pm$
\cite{ZAFM,GHR,HW,HP,SenN=2}.  This is achieved by making the
replacement $\psi^M_\pm\rightarrow \psi^{(J)M}_\pm=J^M_{\pm
  N}\psi^N_{\pm}$ in the $N=1$ supersymmetry transformations
(\ref{N=11})--(\ref{N=13}). The extended supersymmetry transformations
are then given by
\beq
\label{N=21}
\delta_\mp^{(J)}X^M &=&\pm i\epsilon_\mp J^M_{\pm N}\psi_\pm^N
\\
\label{N=22}
\delta_\mp^{(J)}\psi_\pm^M &=&\mp J^M_{\pm N}\del_\pm X^N\epsilon_\mp
\mp i (J^M_{\pm N}\del_L J^N_{\pm K}J^L_{\pm P})\psi_\pm^K
\epsilon_\mp\psi_\pm^P
\\
\label{N=23}
\delta_\mp^{(J)}\psi_\mp^M &=& \mp i
\psi_\mp^N\epsilon_\mp\Omega^{\pm M}_{NK}J^K_{\pm L}\psi_\pm^L .
\eeq
Clearly, for the action (\ref{action}) to be invariant under the
extended supersymmetry transformations, it is sufficient that it is
invariant under the replacement $\psi_\pm\rightarrow\psi^{(J)}_\pm$.
This requires that the metric $G_{MN}$ be Hermitian with respect to
$J^M_{\pm N}$ and that the almost complex structures be covariantly
constant with respect to the torsionful connections
$\Omega^{\pm M}_{NK}$:
\beq
\label{herm}
&J^M_{\pm K}G_{MN}J^N_{\pm L} = G_{KL}&
\\
\label{cov}
&\nabla^\pm_M J^N_{\pm K}\equiv \del_M J^N_{\pm K}+ \Omega^{\pm N}_{ML}
J^L_{\pm K} - J^N_{\pm L}\Omega^{\pm L}_{MK} = 0& .
\eeq
The above conditions ensure that the bilinear fermion terms and the
four-fermion terms in the action (\ref{action}) are separately
invariant under $\psi_\pm\rightarrow\psi^{(J)}_\pm$.
The two sets of supersymmetry transformations,
(\ref{N=11})--(\ref{N=13}) and (\ref{N=21})--(\ref{N=23}), satisfy the
usual $N=2$ algebra, provided the Nijenhuis tensors corresponding to
$J_\pm$ vanish and hence the almost complex structures are integrable.
(For a discussion of the more general case where this is not true, see
\cite{papad}.)  The above discussion can be easily generalized to the
extension of $N=1$ to $N=4$ supersymmetry which requires the existence
of three complex structures satisfying a quaternionic algebra.
It is clear that the existence of an extended supersymmetry on the
worldsheet is related to the geometrical properties of the target
manifold. In particular, eq. (\ref{cov}) restricts the holonomy
of the target manifold. We will see that for the dual theory the
situation is somewhat different. Our results in the following are
independent of the details of the extended supersymmetry and hence the
complex structures we consider could be either associated with an
$N=2$ or $N=4$ supersymmetry.

The issue we want to address now is how T-duality affects the complex
structures and therefore the extended worldsheet supersymmetry. Here,
there are two possibilities: ({\it i}) ${\cal L}_K J=0$ and ({\it
  ii}) ${\cal L}_K J\ne 0$. In the first case, which in our preferred
coordinate system corresponds to $\theta$-independent complex
structures, the answer is known \cite{KIR,H}. In this case, the dual
theory also admits almost complex structures $\wt J_\pm$ given by
\be
\label{Jlocal}
\wt J_{\pm} = Q_{\pm} J_\pm Q^{-1}_\pm ,
\ee
where $Q_\pm$ are defined in (\ref{Qs}). It can be shown that
$\wt J_\pm$ are covariantly constant and integrable \cite{H}.
Therefore, in this case the supersymmetry in the dual model is
realized in the usual manner. The discussion of duality and
supersymmetry in \cite{GHR,RV,KKL} falls in this category.

In the following, we concentrate on the case ${\cal L}_K J_\pm \ne 0$
which corresponds to $\theta$-dependent complex structures. In this
case, it can be easily checked that $\wt J_\pm(\theta, x^i)$, as given
by (\ref{Jlocal}), are no longer covariantly constant. This implies
that the action (\ref{action}) is not invariant under the
corresponding supersymmetry transformations. Examples of this type of
models were first encountered in \cite{Bakas}.  Since duality is a
symmetry of the underlying conformal field theory, one expects that
supersymmetry survives though, in the dual model, it is not realized
in the standard manner.  In fact, the discussion in the previous
section on the canonical approach to duality transformations indicates
that in the dual theory the $\theta$-dependent complex structures are
replaced by non-local objects.  To see this, the first step is to find
the non-local objects which replace the complex structures $J(\theta,
x^i)$ in the dual theory. This can easily be done by requiring the
covariance under duality of the extended supersymmetry transformation
(\ref{N=23}). Let us again denote the objects dual to the complex
structures by $\wt J^M_{\pm N}$.  Then, using (\ref{ct-f}),
(\ref{ct-bp}) and (\ref{dconnec}), we obtain
\be
\label{Jnlocal}
\wt J_{\pm}([\wt\theta, x^i],x^i)
 = Q_{\pm} J_\pm(\theta[\wt\theta, x^i],x^i) Q^{-1}_\pm .
\ee
Here, $\theta[\wt\theta, x^i]$ is the usual notation for the
functional dependence of $\theta$ on $\wt\theta$ and $x^i$ with the
explicit relation given by (\ref{ct-bp}). It is clear that now $\wt
J_\pm$ has a non-local dependence on the coordinates of the dual
target space $\{\wt X^M\}=\{\wt\theta,\,x^i\}$.  For
$\theta$-independent complex structures, (\ref{Jnlocal}) reduces to
(\ref{Jlocal}). Equation (\ref{Jnlocal}) is also in agreement with
the prescription given in \cite{BS} for obtaining the non-local
supersymmetry, except for the inclusion of the
worldsheet fermions in the relation between $\theta$ and $\wt\theta$.

The extended supersymmetry transformations of the dual model can now
be defined in the standard way: as the $N=1$ transformations acting on
$\wt\psi^{(J)}_\pm = \wt J_\pm\wt\psi_\pm$. Since, on shell,
$\wt\psi^{(J)}_\pm$ is a local function of the worldsheet coordinates,
the variation of the dual action under the above transformations can
still be obtained without any complications. However, these
transformations can no longer be written in the form
(\ref{N=21})--(\ref{N=23}) since the derivatives of the non-local
objects $\wt J_\pm$ now have to be properly treated. The derivatives
of $\wt J$ which still make sense are the partial derivatives with
respect to $x^i$ and the derivative with respect to the original
$\theta$ coordinate, {\it i.e.} $\del_i\wt J$ and $\del_\theta \wt
J$. We will therefore express all results in terms of these
derivatives since the ordinary derivative of $\wt J$ with respect to
$\wt\theta$ is not defined (it turns out that a functional $\wt
\theta$-derivative is not the natural object to replace the ordinary
$\theta$-derivative in the dual theory). In particular, using
(\ref{ct-bp}), we can write
\be
\label{dpmwtJ}
\del_\pm\wt J =  \del_i\wt J\,\del_\pm x^i +
\left[(Q^{-1}_\pm)^\theta_L\,\del_\pm\wt X^L +
i\,G^{-1}_{\theta\theta}\,F_\pm\right]\del_\theta\wt J ,
\ee
where $\wt J$ could be either $\wt J_+$ or $\wt J_-$.

We now investigate the conditions under which the action dual to
(\ref{action}) is invariant under the replacement
$\wt\psi_\pm\rightarrow\wt\psi^{(J)}_\pm=\wt J_\pm\wt\psi_\pm$ and,
therefore, under the corresponding extended supersymmetry
transformations. For definiteness, let us focus on the transformations
involving $\wt J_+$. Since $\wt J_+$ is not covariantly constant, the
four-fermion terms and the terms bilinear in $\wt\psi_+$ appearing in
the dual action are not separately invariant under the above
replacement. We first calculate the variation of the four-fermion term
in the dual theory. In (\ref{dualcurv}), we can replace $\psi_+$ by
$\psi^{(J)}_+$ since they both transform in the same way under
duality. This gives the four-fermion term in the dual theory after the
replacement $\wt\psi_+\rightarrow\wt\psi^{(J)}_+$ as
\be
\label{dualcurvJ}
\wt\psi^{(J)M}_+\wt\psi^{(J)N}_+\wt\psi^K_-\wt\psi^L_-
\wt R_{MNKL}(\wt\Omega^-)=
\psi^{(J)M}_+\psi^{(J)N}_+\psi^K_-\psi^L_- R_{MNKL}(\Omega^-)
+2 G^{-1}_{\theta\theta}F^{(J)}_+F_- .
\ee
Now, using the covariant constancy of $J_+$ in $F^{(J)}_+$, we get
\be
F^{(J)}_+ = F_+ + \psi^M_+\,G_{MN}\,J^N_{+L}\,\del_\theta J^L_{+K}\,
\psi^K_+ .
\ee
Notice that $F^{(J)}_+-F_+$ is self-dual. Substituting this back in
(\ref{dualcurvJ}) and using (\ref{dualcurv}), we obtain the variation
of the four-fermion term as
\beq
\wt\psi^{(J)M}_+\,\wt\psi^{(J)N}_+\,\wt\psi^K_-\,\wt\psi^L_-\,
\wt R_{MNKL}(\wt\Omega^-)\!\!&-&\!\!
\wt\psi^M_+\,\wt\psi^N_+\,\wt\psi^K_-\,
\wt\psi^L_-\,\wt R_{MNKL}(\wt\Omega^-)
\nonumber\\
&=&\! 2\,\wt\psi^M_+\,\wt G_{MN}\wt J^N_{+L} \del_\theta\wt J^L_{+K}\,
\wt\psi^K_+ \,G^{-1}_{\theta\theta}\,F_- .
\eeq
To this, we add the variation under the replacement
$\wt\psi_\pm\rightarrow \wt J_\pm \wt\psi_\pm$ of the terms bilinear
in $\psi_+$. Setting the total variation to zero gives us the
condition for the invariance of the dual theory under the non-local
extended supersymmetry as
\be
\label{nlcovws}
\del_-\wt J^M_{+N}
- i G^{-1}_{\theta\theta} \del_{\theta}\wt J^M_{+N} F_- +
\left(\wt \Omega^{+M}_{KL}\,\wt J^L_{+N} - \wt J^M_{+L}\,
\wt \Omega^{+L}_{KN}\right) \del_- \wt X^K = 0 .
\ee
Using (\ref{dpmwtJ}), the above equation (and the similar equation
involving $\wt J_-$) can be written as conditions on $\wt J_\pm$:
\beq
\label{nlcov}
&&\wt G_{\wt\theta\wt\theta}\,\del_\theta \wt J^M_{\pm N}
+\wt \Omega^{\pm M}_{\wt\theta L}\,\wt J^L_{\pm N}
-\wt J^M_{\pm L}\,\wt\Omega^{\pm L}_{\wt\theta N} =0 .
\\
&&\wt\nabla^\pm_i\,\wt J^M_{\pm N} \pm (\wt G\pm \wt B)_{\wt\theta i}
\,\del_\theta \wt J^M_{\pm N} =0
\nonumber
\eeq
The above equations generalize the condition of covariant constancy of
complex structures to the case where the extended supersymmetry of the
dual theory is realized non-locally. Notice that, to avoid any
confusion, in eqs. (\ref{nlcov}) the index in the $\wt\theta$
direction has been explicitly labeled so. In the previous equations,
where there is no risk of confusion, the index $\theta$ was
used to represent either a $\theta$ or a $\wt\theta$ direction.  As a
consistency check, notice that eqs. (\ref{nlcov}) are compatible
with the ones obtained directly from (\ref{Jnlocal}) by assuming that
the $J$ is covariantly constant.

Even when the extended supersymmetry becomes non-local under duality,
the extended superconformal algebra remains unchanged. However, this
algebra is now realized in terms of non-local supercharges and the
representation becomes non-local. Such non-local representations in a
class of conformal field theories were constructed in terms of
parafermions in \cite{Kounnas}. The relevance of these representations
to the behaviour of supersymmetry under duality was discussed in
\cite{BS}. There are several explicit examples known in which a part
of the extended supersymmetry becomes non-local under duality
\cite{Bakas, BS}.  In all of these cases the original theory has an
$N=4$ supersymmetry such that, in each chiral sector, two out of the
three complex structures are $\theta$-dependent. As a result, after a
duality transformation with respect to the $\theta$-isometry, only an
$N=2$ supersymmetry is locally realized. However, it is clear that the
non-localization of extended supersymmetry is a generic feature of
$\theta$-dependent complex structures and is therefore not necessarily
restricted to $N=4$ theories.

As discussed in the previous section (below eq.(\ref{below})), a
special situation arises when $Q_-=1_D$, so that the background fields
$G$ and $B$ are self-dual. It was argued that, in this case, the
non-local dependence on the dual coordinate can be removed by a chiral
shift of $\theta$, which is now a symmetry of the theory. For
backgrounds of this kind, the covariant constancy of $J_-$ implies
that $\del_{\theta}J_-=0$, so that $\wt J_-=J_-$. On the other hand,
$J_+$ can still have a $\theta$-dependence and $Q_+\ne 1_D$.
Therefore, in general, $\wt J_+\ne J_+$, although $\wt J_+$ is still
local and covariantly constant, as can be seen from (\ref{nlcov}).  An
example in which this situation arises is the supersymmetric
$SU(2)\times U(1)$ WZNW model which has extended $N=4$ supersymmetry
on the worldsheet. If we use the usual Euler parametrization for the
$SU(2)$ group, then the resulting theory has a manifest chiral
isometry with respect to which the backgrounds are self-dual. The
coordinate $\theta$ is now conjugate to the $SU(2)$ Cartan generator,
say, $T_3$. The complex structures are defined by their action on the
Lie algebra at the identity and can be extended to the full group
manifold using the left-invariant and the right-invariant one-forms.
Out of the six complex structures (three in each chiral sector), only
two in the left-moving sector are $\theta$-dependent.  Now, it can be
verified that the transformations of the complex structures, as given
by (\ref{Jnlocal}), are consistent with the well-known interpretation
of duality as the automorphism $T_3\rightarrow -T_3$, acting on the
left-moving sector of the worldsheet theory. On the other hand, a
duality with respect to the vector or the axial current leads to the
$SU(2)/U(1) \times U(1)^2$ model in which part of the supersymmetry is
non-locally realized \cite{Kounnas,BS}.

\section{Implications for Target Space Supersymmetry}

In this section we discuss the implications of the non-localization of
extended worldsheet supersymmetry under T-duality for the associated
target space supersymmetry.

In the low-energy limit of superstring theory, the massless background
fields $G_{MN}, B_{MN}$ and $\Phi$, along with their superpartners,
transform under two copies of $N=1$ supersymmetry transformations.
These have their origin in the independent left-moving and
right-moving supersymmetries on the worldsheet. Since the backgrounds
describe a vacuum configuration for the low-energy strings, the theory
describing the fluctuations around these backgrounds will have
unbroken supersymmetry if the backgrounds themselves are invariant.
This can be achieved by setting to zero the fermionic backgrounds,
along with their variations under supersymmetry. In the following, we
explicitly consider one copy of the $N=1$ supersymmetry
transformations and, moreover, set the gravitino background $\Psi_M$
and the dilatino background $\lambda$ to zero.  Let us denote the
space-time supersymmetry transformation parameter by $\eta$ which is a
Majorana-Weyl spinor in $D=10$. Then, setting to zero the variation
under supersymmetry of the fermionic backgrounds gives:
\beq
\label{gravitino}
\delta \Psi_M&=&\del_M\eta +{1\over 4}\left(\omega^{AB}_M -
{1\over 2} H_M^{~~AB}\right) \gamma_{AB}\,\eta =0
\\
\label{dilatino}
\delta\lambda&=&\gamma^M\del_M\,\Phi\,\eta -
{1\over 6} H_{MKN}\gamma^{MKN}\,\eta =0 .
\eeq
Here, $\omega_M^{AB}$ is the spin connection, and the torsion term
added to it has been explicitly exhibited. The indices $A, B$ refer to
the tangent space. Equation (\ref{gravitino}) defines a Killing spinor
on the target space. It is known that if the Killing spinor $\eta$ is
independent of the coordinate $\theta$, then the above equations are
also satisfied in the dual theory \cite{BOK}. In this case, the spinor
$\eta$ does not transform under duality. When $\del_\theta\eta\ne 0$,
the above equations are not satisfied in the dual theory. Also, in
\cite{Bakas} an example was considered and it was argued that eq.
(\ref{dilatino}) is not satisfied after a T-duality transformation.
This suggests that in these cases, the target space supersymmetry of
the dual theory is not realized in the conventional way.

The connection between the extended worldsheet supersymmetry and the
target space supersymmetry, which is characterized by eqs.
(\ref{gravitino}), (\ref{dilatino}), is well known \cite{CHSW,
  Strominger}. The complex structure
associated with the worldsheet supersymmetry can be constructed in
terms of $\eta$ and (up to a constant factor) is given by
\be
\label{ts-ws}
J^M_{+N} = \bar\eta\,\gamma^M_{~N}\,\eta .
\ee
The Killing spinor condition then implies that $\nabla^+_M
J^K_{+N}=0$. From (\ref{ts-ws}), it is evident that if the vielbeins in
$\gamma^M_{~N}=e^M_A\,e_N^B\gamma^A_{~B}$ are chosen to be independent
of $\theta$, then a $\theta$-dependence of $J$ implies a
$\theta$-dependence for $\eta$. The converse, however, is not always
true and $\del_\theta J=0$ does not necessarily imply $\del_\theta
\eta =0$.

Using (\ref{ts-ws}), it is easy to see what happens to the target
space supersymmetry parameter $\eta$ under duality. There are three
possibilities depending on who $\eta$ and the associated complex
structure $J_+$ depend on $\theta$:

\noindent {\it Case 1}\,: Let us first consider the case
$\del_\theta\eta =0$, which implies $\del_\theta J_+ =0$. Under
duality, the transformation of $J_+$ is given by (\ref{Jlocal}). As
discussed below eq. (\ref{dconnec}), the T-dual of the (inverse)
vielbein is not unique and is given by either $\wt e_+=Q_+ e$ or $\wt
e_-=Q_-e$. However, since the two are related by a local Lorentz
transformation, we can choose either of them. If we choose $\wt e_+$
as the dual vielbein, then it is apparent that eq.  (\ref{ts-ws}) is
also valid in the dual theory without transforming the spinor $\eta$.
Since the dual complex structure is covariantly constant, it follows
that $\eta$ is also a Killing spinor for the dual
theory\footnote{Since the same torsion term has been added both to the
  spin connection and to the affine connection, the vielbeins are
  still covariantly constant.}.  For the second copy of target space
supersymmetry, which is associated with $J_-$, it is natural to choose
$\wt e_-$ as the dual vielbein.  This can be re-expressed in terms of
$\wt e_+$, using the local Lorentz transformation relating the two.
{}From the analogue of eq.  (\ref{ts-ws}) for $J_-$, it then follows
that this local Lorentz transformation can be absorbed in a
redefinition of the Killing spinor associated with the second
supersymmetry. Thus, the difference in the transformations of the two
worldsheet chiral sectors under duality results in a redefinition of
the spinors associated with the target space supersymmetry.

\noindent{\it Case 2}\,: Now, we address the issue of target space
supersymmetry when the corresponding extended worldsheet supersymmetry
becomes non-local under duality. As discussed in the previous section,
this situation arises when $\del_\theta J_\pm \ne 0$ and hence
$\del_\theta \eta\ne 0$. In this case, the non-local object $\wt J_+$,
which is dual to the complex structure $J_+$, is given by eq.
(\ref{Jnlocal}). For the dual vielbein, we again choose $\wt e_+$.
Equation (\ref{ts-ws}) then implies that in the dual theory there
exists a non-local spinor $\wt\eta$ given by
\be
\label{deta}
\wt \eta ([\wt\theta, x],x) = \eta (\theta[\wt\theta,x],x) .
\ee
The analogue of the Killing spinor condition (\ref{gravitino}) for
$\wt\eta$ can be obtained by substituting $\wt J^M_{+N}=
\bar{\wt\eta}\wt\gamma^M_{~N}\wt\eta$ in eq. (\ref{nlcov}) which is
a generalization of the covariant constancy condition for $\wt J_+$.
Corresponding to the $\theta$-component and the $i$-components of the
Killing spinor condition (\ref{gravitino}), we obtain the following two
equations in the dual theory:
\beq
\label{nlgravitino}
&&\wt G_{\wt\theta\wt\theta}\,\del_\theta\,\wt\eta +
{1 \over 4}\left(\,\wt\omega^{AB}_{\wt\theta} - {1\over 2}
\wt H_{\wt\theta}^{~~AB}\,\right)\gamma_{AB}\,\wt\eta =0
\\
&&\del_i\wt\eta +{1\over 4} \left(\,\wt\omega^{AB}_i -
{1\over 2}\wt H_i^{~~AB}\,\right) \gamma_{AB}\,\wt\eta
+\,(\wt G+\wt B)_{\wt\theta i}\,
\del_\theta\,\wt\eta = 0 .
\nonumber
\eeq
The correctness of the above equations can also be checked directly by
using the relation between $\eta$ and $\wt\eta$ in (\ref{gravitino}).
In the dual theory, we have not written down the equation that
corresponds to the vanishing of the dilatino variation
(\ref{dilatino}). This equation can be obtained by substituting the dual
variables in (\ref{dilatino}).  Equations (\ref{nlgravitino}) reduce
to the usual Killing spinor equation when either $\del_\theta \wt\eta
=0$ or $Q_- =1$. The later case, as discussed in the previous
sections, corresponds to a self-dual theory with a chirally conserved
current.

\noindent{\it Case 3}\,: The only other possibility is when the target
space spinor $\eta$ depends on the coordinate $\theta$ in such a way
that the $\theta$-dependences on the right hand side of (\ref{ts-ws})
cancel out, giving rise to a $\theta$-independent $J$. In this case,
in the dual theory, the extended worldsheet supersymmetry is locally
realized while the associated target space supersymmetry has a
non-local realization. It turns out that the model considered in
\cite{Bakas} and \cite{BOK}, in connection with the apparent
violations of supersymmetry under T-duality, falls in this class. In
the following we briefly describe this model in order to clarify its
behaviour based on our approach. Consider the four-dimensional flat
Euclidean space $\{X^1, X^2, X^3, X^4\}$. This space admits three
complex structures $J^a, \,a=1,2,3$, which satisfy a quaternionic
algebra and the corresponding theory has $N=4$ supersymmetry. When the
metric is flat, one of the complex structures (say $J^3$) is in the
canonical form.  Now, let us choose polar coordinates in the
$X^1X^2$-plane: $\{X^1,X^2\}\rightarrow \{r,\theta\}$. Two of the
complex structures, $J^1$ and $J^2$, develop a $\theta$-dependence
while $J^3$ and the metric are independent of $\theta$. To keep the
vielbeins also $\theta$-independent, we have to perform a
$\theta$-dependent local Lorentz transformation. As is evident from
(\ref{ts-ws}), this Lorentz transformation can be absorbed in the
target space spinors associated with $J^a$, thus making them
$\theta$-dependent. Now, under a T-duality transformation with respect
to $\theta$, all target space and worldsheet supersymmetries become
non-local except for the worldsheet supersymmetry associated with
$J^3$.

In some cases, such as the $SU(2)\times U(1)$ WZNW model, the
non-local nature of the target space supersymmetry can also be
interpreted in a somewhat different way. In this model there is a
natural choice for the vielbeins in terms of the left-invariant or
right-invariant one-forms. In this case, although the metric is still
$\theta$-independent, the vielbeins are not. The full
$\theta$-dependence of the complex structure in (\ref{ts-ws}) is then
contained in the vielbein and not in the spinor $\eta$. As a result,
in the dual theory, $\wt\eta$ is local but the vielbein transforms
into a non-local object. In this scenario, the dual target space
possesses a supersymmetry that is defined not in a standard local
Lorentz frame, but in a frame connected to the latter by a non-local
rotation.

Since duality is a symmetry, one expects that the dual target space
theory also admits some kind of supersymmetry with the non-local
spinor $\wt\eta$ as the transformation parameter. Though the explicit
form of this transformation is not known, in the following, we will
argue that they are expected to be very different from the standard
target space supersymmetry transformations. To see this, note that
since the dual backgrounds are expressed in terms of the original
ones, it follows that if the fermionic backgrounds in the original
theory are set to zero, they will remain zero in the dual theory. This
is consistent with the invariance of the dual backgrounds under the
non-local supersymmetry. It is then reasonable to expect, in analogy
with the local case, that the gravitino variation in the dual theory
is proportional to the left-hand side of (\ref{nlgravitino}). The form
of this transformation is clearly different from the usual
supersymmetry transformation of the gravitino given by
(\ref{gravitino}). In particular, due to its non-locality, this
transformation makes sense only when the coordinates are restricted to
the string worldsheet, and not at a generic space-time point.  The
existence of this transformation in the dual theory would not have
been evident in the absence of the duality relation.  Since the
background fields are invariant under supersymmetry, the non-locality
does not show up as long as we are looking only at the vacuum
configurations. However, one expects that the modification of the
supersymmetry transformations should have important consequences for
the spectrum of fluctuations around these backgrounds which are the
relevant quantum fields for the low-energy theory. In particular, it
is unlikely that the dual theories are equivalent as quantum field
theories.

As string theories, the equivalence under duality is a consequence of
the existence of both momentum and winding modes associated with the
compact coordinate $\theta$. It is well known that under duality these
modes are interchanged: The conserved momentum $P_\theta$ and the
winding number $L_\theta$ associated with the compact coordinate
$\theta$ (with non-trivial $\pi_1$) are given by $P_\theta =
\int_0^{2\pi} d\sigma p_\theta$ and $L_\theta=\theta(\sigma=2\pi)-
\theta(\sigma=0)$. Then, from the canonical transformations
(\ref{ct-b}), it follows that $\wt P_\theta=-L_\theta$ and $\wt
L_\theta=-P_\theta$.  Since the momentum and winding modes are
associated with the worldsheet coordinates $\tau$ and $\sigma$,
respectively, their interchange under duality is the origin of the
non-local relationship between $\theta$ and $\wt\theta$. This can be
easily see when the backgrounds are flat and one can write $\theta=
\theta_L + \theta_R$ whereas, $\wt\theta = \theta_L-\theta_R = \int
d\sigma^+ \del_+\theta - \int d\sigma_- \del_-\theta$.  As for the
behaviour of supersymmetry, note that the parameter $\eta(\theta,x)$
is sensitive to the string momentum and winding modes associated with
$\theta$. The non-locality in the dual theory arises due to the fact
that the the momentum and winding modes of the dual string enter
$\wt\eta$ not through $\wt\theta$ (which would have resulted in a
local spinor $\wt\eta (\wt\theta)$), but through the original
coordinate $\theta$.

\section{Summary and Discussions}

In this section, we first summarize our results and then briefly
discuss their generalization to $O(d,d)$ deformations. At the end, we
discuss an apparent violation of worldsheet supersymmetry under
S-duality transformations.

We have addressed the issue of the non-locality of extended worldsheet
supersymmetry and the associated target space supersymmetry under
T-duality transformations. This happens when the complex structure $J$
associated with the extended supersymmetry has a dependence on the
coordinate (say $\theta$) with respect to which duality is performed.
To study this issue systematically, we first generalized the
implementation of a T-duality by a canonical transformation from the
bosonic to the supersymmetric non-linear \s-model.  Using this, we
obtained the non-local object $\wt J$ which, in the dual theory,
replaces the $\theta$-dependent complex structure $J$. Similar to the
complex structure, this non-local object defines the extended
supersymmetry of the dual model in terms of its default $N=1$
supersymmetry. The extended supersymmetry of the dual model is thus
realized non-locally. The non-locality is only in terms of the target
space coordinates and, on shell, the theory is local in terms of the
worldsheet coordinates.  The invariance of the dual action under the
non-local supersymmetry imposes some restrictions on $\wt J$, which
are analogous to the covariant constancy condition for the complex
structure in the usual realization of the supersymmetry. We then used
the relation between the extended worldsheet supersymmetry and the
target space supersymmetry of the original theory to argue that the
dual target space admits supersymmetry transformations with a
non-local spinor parameter. We also obtained the analogue of the
Killing spinor equation which this non-local spinor satisfies. The
analysis suggests that the action of the non-local supersymmetry on
the backgrounds is different from the action of the standard target
space supersymmetry.  Thus, the supersymmetry of the dual theory has a
non-standard realization on the spectrum of fluctuations around these
backgrounds and the two theories are not equivalent as field theories.
The equivalence as string theories is a consequence of the presence of
winding (or winding like) modes in the string spectrum. The emergence
of non-local effects in the low-energy theory is related to the
momentum-winding interchange under T-duality.
When the backgrounds are self-dual, the isometry with respect
to which duality is performed becomes chiral.  It was shown that in
this case the non-locality can be removed by a chiral shift of the
coordinate.  In these situations, as expected, both the worldsheet and
the target space supersymmetries remain local.

The mechanism by which the non-locality appears in the theory is not
specific to duality transformations. On the contrary, it is common to
all non-trivial $O(d,d)$ transformations which contain the T-duality
transformations as a discrete subgroup \cite{GRV,MV}: Let $\theta^m$
denote the $d$ coordinates on which the background fields do not
depend, and let $p_m$ denote the corresponding conjugate momenta. If we
define a $2d$ -dimensional vector $Z$ as $Z^T=(-\theta^{'T}, p^T)$,
then an $O(d,d)$ transformation can be implemented by the canonical
transformation $\wt Z = \Omega Z$, where $\Omega\in O(d,d)$. Notice
that these canonical transformations do not in general lead to
equivalent quantum theories but rather correspond to the deformations
of the original theory \cite{def}. The transformation of the complex
structures is again given by an equation similar to (\ref{Jnlocal}),
with the difference that now $Q_\pm$ are the general matrices given in
\cite{H}. This shows that theories with extended supersymmetry, which
have $\theta^m$-dependent complex structures and also admit $O(d,d)$
deformations, actually correspond to very special points in a theory
space where, generically, the extended supersymmetry has a non-local
realization. An example of this is the space of deformations of the
$SU(2)\times U(1)$ WZNW model with $N=4$ worldsheet supersymmetry.

Recent work has revealed a close connection between T-duality and
S-duality transformations, notably the fact that in some theories
their roles get interchanged (see \cite{ST} and references therein).
It is known that S-duality acts as an R-symmetry on the target space
spinors \cite{O}. However, there exists an example where the extended
worldsheet supersymmetry is not preserved under S-duality (to be more
precise, a one-parameter family of $SL(2,R)$ transformations).  This
example, which involves an intertwining of S- and T-dualities, was
considered in \cite{Bakas} in connection with the apparent
supersymmetry violations of T-duality and was also described in the
previous section (under Case 3). There we saw that in this model, after
a T-duality, the target space supersymmetries are non-locally realized
though we are still left with one locally realized worldsheet
supersymmetry.  A combination of S-T transformations on this
background again leads to a pure gravitational background with a
metric that is Ricci-flat but not hyper-K\"{a}hler. This means that
the metric could not be K\"{a}hler and therefore does not admit a
covariantly constant complex structure.  This in turn shows that the
surviving $N=2$ supersymmetry is no longer manifest after the S- and
T-duality transformations.  There are two possibilities for this to
happen. One possibility is that the surviving complex structure
develops a $\theta$-dependence as a result of the S-duality and
therefore becomes non-local after the final T-duality. The other
possibility is that the complex structure does not survive the
S-duality. An explicit calculation shows that it is the second
possibility that actually occurs. We therefore have a situation where
S-duality destroys a complex structure. Note that there is no
contradiction with the space-time supersymmetry since there are no
locally realized target space supersymmetries.  Unlike the case of
T-duality, it is not clear what the origin of this phenomenon is and
what happens to the supersymmetry associated with this complex
structure.  \vskip .5cm
\noindent{\Large{\bf Acknowledgements}}
\vskip .5cm
\noindent I would like to thank I. Bakas, E. Kiritsis,
L. \'{A}lvarez-Gaum\'{e}, A. Sen and K. Sfetsos for useful
discussions.
\vskip .5cm
\noindent{\Large{\bf Note Added}}
\vskip .5cm
\noindent The problem of duality and supersymmetry has recently been
studied in \cite{AEB} from a different point of view.

\end{document}